\documentclass[aoas,MSNbibl,nameyear,dvips]{arximspdf}
\usepackage{multirow}
\usepackage{dcolumn}
\usepackage{graphicx}

\doi{10.1214/11-AOAS496}
\volume{6}
\issue{1}
\pubyear{2012}
\firstpage{356}
\lastpage{382}

\makeatletter

\setattribute{abstract}   {width}  {300pt}

\newcolumntype{d}[1]{D{.}{.}{#1}}

\makeatother

\begin{document}
\begin{frontmatter}

\title{A Bayesian measurement error model for two-channel cell-based RNAi data with~replicates\thanksref{T1}}
\runtitle{A Bayesian model for RNAi data with replicates}

\thankstext{T1}{Supported in part by NRPGM Grants NSC 95-3112-B001-019,
NSC 96-3112-B001-003, NSC 96-3112-B001-012, NSC 97-3112-B001-001,
NSC-98-3112-B-400-009 and NSC-98-3112-B-400-010.}

\begin{aug}
\author[A]{\fnms{Chung-Hsing} \snm{Chen}},
\author[B]{\fnms{Wen-Chi} \snm{Su}},
\author[A]{\fnms{Chih-Yu} \snm{Chen}},
\author[B]{\fnms{Jing-Ying}~\snm{Huang}},
\author[A]{\fnms{Fang-Yu} \snm{Tsai}},
\author[A]{\fnms{Wen-Chang} \snm{Wang}},
\author[A]{\fnms{Chao~A.}~\snm{Hsiung}},
\author[B]{\fnms{King-Song} \snm{Jeng}}
\and
\author[A]{\fnms{I-Shou} \snm{Chang}\corref{}\ead[label=e1]{ischang@nhri.org.tw}}

\runauthor{C.-H. Chen et al.}
\affiliation{National Health Research Institutes, Academia Sinica,
National Health Research Institutes, Academia Sinica,
National Health Research Institutes, National Health Research
Institutes, National Health Research Institutes, Academia Sinica,
and~National Health Research Institutes}
\address[A]{C.-H. Chen\\
C.-Y. Chen\\
F.-Y. Tsai\\
W.-C. Wang\\
C. A. Hsiung\\
I-S. Chang\\
Division of Biostatistics\\
\quad and Bioinformatics\\
National Health Research Institutes\\
Miaoli, Taiwan\\
\printead{e1}} 
\address[B]{W.-C. Su\\
J.-Y. Huang\\
K.-S. Jeng\\
Institute of Molecular Biology\hspace*{27.5pt}\\
Academia Sinica\\
Taipei, Taiwan}
%
%
\end{aug}

\received{\smonth{3} \syear{2010}}
\revised{\smonth{7} \syear{2011}}

%

\begin{abstract}
RNA interference (RNAi) is an endogenous cellular process in which
small double-stranded RNAs lead to the destruction of mRNAs with
complementary nucleoside sequence. With the production of RNAi
libraries, large-scale RNAi screening in human cells can be conducted
to identify unknown genes involved in a biological pathway. One
challenge researchers face is how to deal with the multiple testing
issue and the related false positive rate (FDR) and false negative rate
(FNR). This paper proposes a Bayesian hierarchical measurement error
model for the analysis of data from a two-channel RNAi high-throughput
experiment with replicates, in which both the activity of a~particular
biological pathway and cell viability are monitored and the goal is to
identify short hair-pin RNAs (shRNAs) that affect the pathway activity
without affecting cell activity. Simulation studies demonstrate the
flexibility and robustness of the Bayesian method and the benefits of
having replicates in the experiment. This method is illustrated through
analyzing the data from a RNAi high-throughput screening that searches
for cellular factors affecting HCV replication without affecting cell
viability; comparisons of the results from this HCV study and some of
those reported in the literature are included.
\end{abstract}

%

\end{frontmatter}

\section{Introduction}\vspace*{-4pt}
\label{INTRODUCTION}
\subsection{RNA interference high-throughput screening and the
motivating example}\label{RNAinterference}

$\!\!$RNA interference (RNAi) is a conserved biological pathway by~which
messenger RNAs are targeted for degradation by double stranded RNA of
identical sequence and, thus, it silences gene expression on the level
of individual transcripts [\citet{C2}, \citet{F1}]. For example, in mammalian cells, small interfering RNAs
(siRNAs) are effective in silencing target mRNAs [\citet{E1}, \citet{E2}, \citet{H1}]. Initially,
it was used to knockdown the function of individual genes of interest;
with the production of RNAi libraries, it is possible for the current
technology to silence most of the genes in the genome and conduct
genome-wide loss-of-function screening so as to identify previously
unknown genes involved in a biological pathway; it provides a
systematic analysis of the genome. The purpose of RNAi high-throughput
screening (HTS) is to identify a set of siRNAs affecting the cellular
phenotype of interest; in the HTS literature, this is referred to as
hit selection. For example, RNAi technology has opened up the field of
genomic scale cell-based screening to the study of viral-host
interactions [\citet{C4}] and several RNAi HTS have been conducted
to identify cellular factors required for various viral infections.

As \citet{B3} and \citet{E1} pointed out,
while RNAi HTS is promising and has generated various significant
technical advances and scientific findings, there are still challenges
regarding high-throughput assay development and data analysis. In
particular, \citet{C4} and \citet{T1} remarked that in the
studies of viral-host interactions, while RNAi HTS has led to the
discovery of hundreds of new factors and increased our knowledge of the
host factors that impact viral infection and highlighted the cellular
pathways at play, there is a surprising lack of concordance between the
results of seemingly similar screens. Among issues to be considered
when comparing these studies, one recognizes the need to standardize
the statistics methods and to address the false positive and false
negative rates inherent to high-throughput siRNA screens.

We note that RNAi HTS refers to a wide range of different experiments.
To determine the assay appropriate for the biological process to be
investigated and to choose or develop statistical methods suitable for
data resulting from the experiments are of great importance; see
\citet{B3}. More specifically, this paper treats the
simple situation that uses cell-based homogeneous assay, in which the
phenotypes of many cells are averaged across each well in a microtitre
plate. In particular, we are interested in data generated from a
two-channel cell-based RNAi HTS experiment where the phenotype of a
pathway-specific reporter gene and that of a constitutive reporter are
measured; such experimental setups are typically used in screens for
signaling pathway components; see page R66.5 of
\citet{B4}. Typical examples include RNAi HTS experiments designed
for the identification of cellular factors required for viral
infections, in which cell viability is measured and used to account for
the effect of unequal cell numbers in different wells on the
measurements of virus RNA replication.

This experimental setup makes it possible to distinguish changes in the
readout caused by depletion of the specific pathway components and
changes incurred by the changes in the overall cell number. Based
on
the plot for a~two-channel RNAi HTS experiment, \citet{B4}
suggest, for a given siRNA, the ratio of the log of the readout of the
specific pathway to the log of the overall cell number is used as a
summary of the effect of the siRNA on the specific pathway. In the
analysis of data from RNAi HTS to identify host factors involved in
\mbox{HIV-1} replication, \citet{Boretal10} also carried out log
transformation of the raw data, including both the measurement of the
pathway activity and the cell count in each well, and applied a robust
locally weighted regression [\citet{C5}] to smooth the scatterplot
of the log transformed data. In a sense, this seems to be an extension
and improvement of the suggestion of \citet{B4}.

To make a more systematic use of the data, we propose a Bayesian
regression model that treats cell viability as a covariate and the
specific pathway activity measurement as the response variable so that
the pathway activity change can be studied in terms of the regression
coefficient and the issues of false positive and false negative rates
can be handled in the Bayesian framework.

Statistical methods for cell-based RNAi HTS have been introduced and
reviewed by \citet{B4}, \citet{Z2},
\citet{Z1}, \citet{M1}
and \citet{B1}, among
others. Most of them deal with one-channel screening and use
descriptive quantities like fold change or simple statistics like
$Z$-score or $t$-test to identify a set of siRNAs that inhibit or activate
defined cellular phenotypes. The only exceptions are \citet{Z1}, who proposed a Bayesian method for one-channel screening
experiments to handle false positive and false negative rates, and
\citet{B4}, who included some discussions on
two-channel experiments.

Another feature of our experiment is that for each siRNA, the
experiment is replicated in $2J$ wells, where $J$ of them measure cell
viability and the other~$J$ measure the specific pathway activity. We
note that \citet{T1} is an example of two-channel RNAi HTS with
$J=2$ replicates, with pooled siRNAs though, and that the dual channel
experiment mentioned in \citet{B4} also has $J=2$
replicates. Our method will make use of these replicates.

Our method is motivated by the following RNAi HTS that searches for the
cellular factors that affect HCV replication without affecting cell
viability, referred to as the HCV study in this paper. HCV (hepatitis C
virus) is a small, enveloped, positive-sense single-stranded RNA virus
of the family Flaviviridae and the cause of hepatitis C in humans; see
\citet{R2}. It is estimated that hepatitis C has infected
nearly $200$ million people worldwide, and is now infecting $3$ to $4$
million people per year (WHO, see the URL below). It is currently a
leading cause of cirrhosis, a common cause of hepatocellular carcinoma.
As a result of these conditions, it is the leading reason for liver
transplantation in, for example, the United States (NIDA, see the URL below).

The experiment is carried out as follows. A Huh-7-derived HCV-Luc cell
line that contains the genome of the HCV and harbors the luciferase
gene as a reporter is employed as a cell-based system for RNA
interference screening. We have shown that luciferase activity
correlates well with the HCV replication (data not shown). The HCV-Luc
cells are transduced by the lentiviruses each carrying specific
short-hairpin RNA (shRNA) and puromycin-resistant gene. In these
circumatances, lentivirus-transduced cells survive under puromycin
selection. HCV RNA replication in the well can now be measured in terms
of the expression of luciferase in HCV-Luc and cell viability in the
well can be measured by a colorimetric method which monitors the level
of dehydrogenase in viable cells.

A VSV-G pseudotyped lentivirus-based RNAi library targeting a whole
panel of human kinases and phosphatases was employed [\citet{M3}]
in this study. This library includes $6\mbox{,}390$ shRNAs designed to target
1,187 genes. HCV-Luc cells are seeded in 96-well plates and each well
is transduced with one of these $6\mbox{,}390$ shRNAs. There are in total $71$
plates and each plate has about $6$ wells for the controls. There are
three types of controls: spike negative (SN), no transduction no
puromycin selection (NTNP), and no tranduction with puromycin selection
(NTWP). In SN control wells, the nonhuman gene targeting shLacZ is used
instead of the above-mentioned shRNAs; data from these wells serve as
the RNAi machinery competition controls. In NTNP control wells, cells
are not transduced by lentiviruses nor subjected to puromycin
selection. In NTWP control wells, cells are not transduced by
lentiviruses but subjected to puromycin selection. There are in total
$142$ SN wells with most plates having $2$ SN wells in each plate,
$217$ NTNP wells with most plates having $3$ NTNP wells in each plate
and $67$ NTWP wells with most plates having one NTWP well in each
plate. It is believed that in both SN and NTNP wells, we should observe
normal HCV RNA replication and in NTWP wells, we should observe little
HCV RNA replication and minimal cell viability.

The experiment described in the previous two paragraphs is replicated
four times in a plate-by-plate manner; specifically, for each plate,
there are three replicated plates in the sense that wells at the same
position in these four plates are transduced by the same shRNA. We call
this platewise replication. Among these four plates, we use two of them
to measure cell viability and the other two to report luciferase
activity as a surrogate for HCV RNA replication assay.

%
\begin{figure}
\begin{tabular}{@{}cc@{}}

\includegraphics{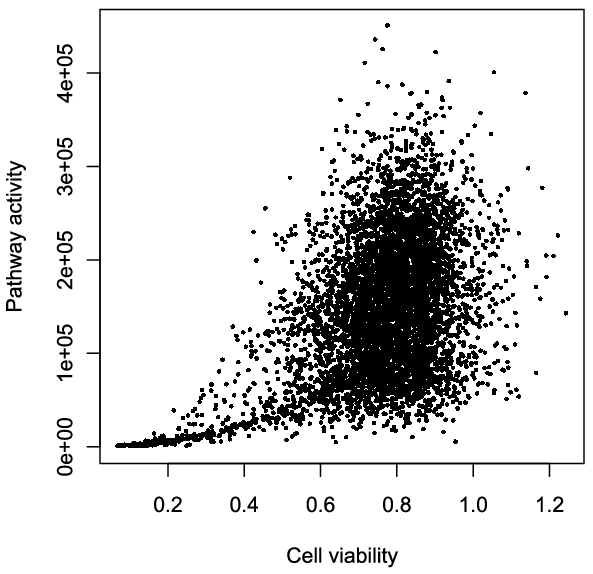}
  & \includegraphics{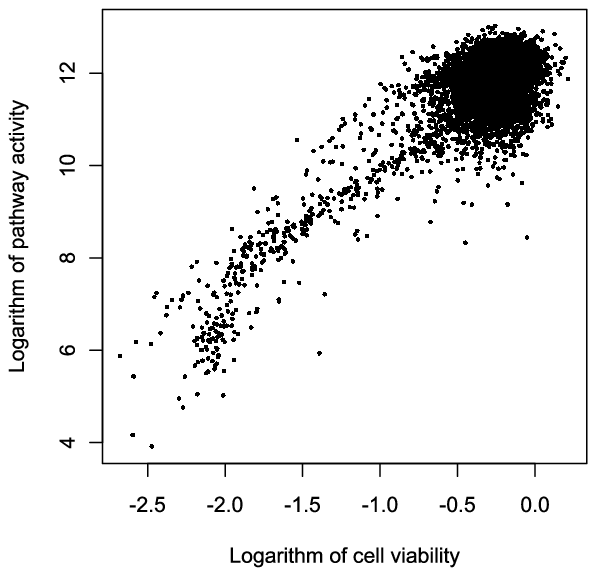}
\\
(a) & (b)
\end{tabular}
\caption{Plots for the HCV study data after preprocessing. There are in
total $6\mbox{,}130$ data points with the $419$ controls excluded.
\textup{(a)} On the original measurement scale;
\textup{(b)} on the log transformed
scale.} \label{RawAndLogDataPlot}
\end{figure}

\subsection{A log-linear measurement error model}\label{Alog-linear}
Let $\tilde{x}_{ij}$ denote the measurement of the cell viability in
the $j$th well of the $i$th shRNA. Let $\tilde{y}_{ij}$ denote the
measurement of the pathway activity in the $(J+j)$th well of the $i$th
shRNA. Here $i=1,\ldots,I$ and $j=1,\ldots, J$ with $J>1$. In practice,
both $\tilde{x}_{ij}$ and $\tilde{y}_{ij}$ refer to the measurements
undergone in certain data preprocessing steps so as to have reduced
systematic bias. Figure \ref{RawAndLogDataPlot} gives the data plot;
Figure \ref{RawAndLogDataPlot}(a) is a plot of the data from the above HCV study with~$\tilde
{x}_{ij}$ and~$\tilde{y}_{ij}$, respectively, the horizontal and
vertical coordinates for $I=6\mbox{,}549$ and $j=1$; Figure~\ref{RawAndLogDataPlot}(b) is the
corresponding plot for $\log\tilde{x}_{ij}$ and $\log\tilde{y}_{ij}$.
Visual examination of Figure~\ref{RawAndLogDataPlot}(a),
(b) suggests that a~linear
model on the logarithm transformed data is considered. In fact, the
Additional data file~4 in \citet{B4} also suggests
the consideration of logarithm transformed data in a~two-channel
experiment based on their data. Another reason for this logarithm
transformation is that based on Q--Q plots, both $\log\tilde
{x}_{i1}-\log\tilde{x}_{i2}$ and $\log\tilde{y}_{i1}-\log\tilde
{y}_{i2}$ are, respectively, much closer to normal distributions than
$\tilde{x}_{i1}-\tilde{x}_{i2}$ and $\tilde{y}_{i1}-\tilde{y}_{i2}$. In
fact, we found that logarithm transformation is very close to those
suggested by the Box--Cox transformation [\citet{B5}] technique.

Motivated by the above exploration, we assume that there exist $\mu_i$
and~$\nu_i$ such that $\log(\tilde{x}_{i1})-\mu_i$ and~$\log(\tilde
{x}_{i2})-\mu_i$ are independent with mean zero and likewise $\log
(\tilde{y}_{i1})-\nu_i$ and $\log(\tilde{y}_{i2})-\nu_i$ are also
independent with mean zero.\looseness=-1

We note that this formulation takes advantage of the replication in the
experiment design and $e^{\mu_i}$ and $e^{\nu_i}$ represent,
respectively, the expected cell viability and pathway activity when
transduced by the $i$th shRNA. For the $i$th shRNA, denote by $\gamma
_i=0$ if it incurs no change on the pathway activity, and by $\gamma
_i=1$ if it does; in case $\gamma_i=1$, the magnitude of the change is
represented by $\beta_i$. $\gamma_i$ is usually called the model index
parameter. The plot in Figure \ref{RawAndLogDataPlot}(b) suggests a linear relation between $\nu
_i$ and $\mu_i$ and the following simple relation
%
\begin{equation}\label{eqsimpleRelation}
\nu_i=\alpha_0+ \gamma_i\beta_i+ \alpha_1\mu_i
\end{equation}
is assumed in this paper, which seems to be one of the simplest that
can be reasonably imposed on the relation between pathway activity and
cell viability. In the context of the HCV study, (\ref
{eqsimpleRelation}) is a simple way to account for the effect of cell
numbers on HCV replication. We note that the HCV infects the host cell
and replicates in it, therefore, the number of cells in a well can
affect the total number of HCV virion in that well; the purpose of
measuring cell viability in each well is to account for this effect on
HCV replication. We regard (\ref{eqsimpleRelation}) as a statistical
formulation to refine and implement the idea of \citet{B4} and \citet{Boretal10} mentioned in
Section \ref{RNAinterference}. In the original measurement scale of
Figure \ref{RawAndLogDataPlot}(a), it says the pathway activity $e^{\nu_i}$ is a monomial of
the cell viability $e^{\mu_i}$ with coefficients involving a baseline
constant $\alpha_0$, activity change indicator $\gamma_i$, and activity
change coefficient $\beta_i$:
\[
e^{\nu_i} =e^{\alpha_0 + \gamma_i \beta_i} \cdot e^{\alpha_1 \mu_i}.
\]

To complete the likelihood specification, we consider the robust and
flexible model which assumes $t$-distributions for each shRNA:
%
\begin{eqnarray}\label{eqregression}
x_{ij} &\equiv& \log(\tilde{x}_{ij})=\mu_i+\varepsilon_{x_{ij}} / \sqrt
{\omega_{x_{ij}}},\nonumber\\[-8pt]\\[-8pt]
y_{ij} &\equiv& \log(\tilde{y}_{ij})=\nu_i+\varepsilon_{y_{ij}} /
\sqrt{\omega_{y_{ij}}}.\nonumber
\end{eqnarray}
Here $\varepsilon_{x_{ij}}$, $\varepsilon_{y_{ij}}$, $\omega_{x_{ij}}$, and
$\omega_{y_{ij}}$ are independent, $\varepsilon_{x_{ij}}$ and
$\varepsilon
_{y_{ij}}$ have normal distributions $\mathcal{N}(0,\sigma_{x_i}^2)$
and $\mathcal{N}(0,\sigma_{y_i}^2)$, respectively, and $\omega
_{x_{ij}}$ and $\omega_{y_{ij}}$ have gamma distributions $\mathcal
{G}a(d_x/2,2/d_x)$ and $\mathcal{G}a(d_y/2,2/d_y)$, respectively. We
note that the error terms $\varepsilon_{x_{ij}} / \sqrt{\omega_{x_{ij}}}$
and $\varepsilon_{y_{ij}} / \sqrt{\omega_{y_{ij}}}$ have $t$-distributions
with degrees of freedom, respectively, $d_x$ and $d_y$ and scale
parameters, respectively, $\sigma_{x_i}^2$ and $\sigma_{y_i}^2$.

Discussions on error terms of the form (\ref{eqregression}) and their
advantages appeared in the gene expression literature; see, for
example, \citet{G4}, \citet{L2} and \citet{L3}. The idea is that a~hierar\-chical $t$-formulation makes
the model more robust to outliers than the usual Gaussian model and an
exchangeable prior for the variance allows each shRNA to have a
different variance and hence makes it more flexible.

Let $\gamma=(\gamma_1,\ldots,\gamma_I)$, $\beta=(\beta_1,\ldots, \beta_I)$,
$\mu=(\mu_1,\ldots,\mu_I)$, $\sigma_x^2=(\sigma_{x_1}^2,\ldots,\sigma
_{x_I}^2)$, $\sigma_y^2=(\sigma_{y_1}^2,\ldots,\sigma_{y_I}^2)$, $\omega
_x=(\omega_{x_1},\ldots,\omega_{x_I})\equiv((\omega_{x_{11}},\ldots,\omega
_{x_{1J}}),\ldots,(\omega_{x_{I1}},\ldots,\break\omega_{x_{IJ}}))$, $\omega
_y=(\omega_{y_1},\ldots,\omega_{y_I})\equiv((\omega_{y_{11}},\ldots,\omega
_{y_{1J}}),\ldots,(\omega_{y_{I1}},\ldots,\omega_{y_{IJ}}))$,
$x=(x_1,\break\ldots,x_I)\equiv((x_{11},\ldots,x_{1J}),\ldots,(x_{I1},\ldots,x_{IJ}))$,
and $y=(y_1,\ldots,y_I)\equiv((y_{11},\ldots,\break y_{1J}),\ldots,(y_{I1},\ldots,y_{IJ}))$.
To study the likelihood, it seems easier to consider the conditional density of
$(x,y)$ given $\omega_x$ and $\omega_y$ or the joint density of~$(x,y)$, $\omega_x$, and~$\omega_y$.
In fact, the former has the closed form:
\begin{eqnarray*}
&&
f(x,y| \gamma, \beta, \mu, \alpha_0, \alpha_1, \sigma_x^2, \sigma
_y^2, \omega_x, \omega_y) \\
&&\qquad=\prod_{i=1}^{I} \prod_{j=1}^{J}
\sqrt{\frac{\omega_{x_{ij}}}{2\pi\sigma_{x_i}^2}} \exp\biggl\{ -
\frac{1}{2} \biggl( \frac{x_{ij}-\mu_i}{\sigma_{x_i} /
\sqrt{\omega_{x_{ij}}}} \biggr)^2 \biggr\}\\
&&\qquad\hphantom{=\prod_{i=1}^{I} \prod_{j=1}^{J}}
{}\times
\sqrt{\frac{\omega_{y_{ij}}}{2\pi\sigma_{y_i}^2}} \exp\biggl\{ -
\frac{1}{2} \biggl( \frac{y_{ij} - \alpha_0 - \gamma_i \beta_i -
\alpha_1\mu_i}{\sigma_{y_i} / \sqrt{\omega_{y_{ij}}}} \biggr)^2
\biggr\}.
\end{eqnarray*}
The latter is equal to
%
\begin{eqnarray}\label{eqlikelihood}
&& g(x, y, \omega_x, \omega_y| \gamma, \beta, \mu, \alpha_0, \alpha_1,
\sigma_x^2, \sigma_y^2, d_x, d_y)\nonumber\\
&&\qquad=f(x,y| \gamma, \beta, \mu, \alpha_0,
\alpha_1, \sigma_x^2, \sigma_y^2, \omega_x, \omega_y)
\nonumber\\[-8pt]\\[-8pt]
&&\qquad\quad{}\times \prod_{i=1}^{I} \prod_{j=1}^{J} \frac{(d_x/2)^{d_x/2}}{\Gamma
(d_x/2)} \omega_{x_{ij}}^{d_x/2-1} \exp\biggl\{ -\frac{d_x}{2} \omega
_{x_{ij}} \biggr\} \nonumber\\
&&\qquad\quad\hphantom{{}\times \prod_{i=1}^{I} \prod_{j=1}^{J} }
{}\times\frac{(d_y/2)^{d_y/2}}{\Gamma(d_y/2)} \omega
_{y_{ij}}^{d_y/2-1} \exp\biggl\{ -\frac{d_y}{2} \omega_{y_{ij}}
\biggr\}.\nonumber
\end{eqnarray}
These expressions are useful in the implementation of Bayesian analysis.

\subsection{Organization of this paper}\label{ABayesian}
Based on the log-linear measurement error model that allows error terms
having shRNA specific $t$-distributions, we will take a Bayesian
hierarchical approach to analyze the data from the HCV study, which is
a typical viral-host interaction study using cell-based RNAi HTS. Our
purpose is to propose shRNA lists and associated false positive rates
so that experimental scientists can decide whether to follow up with
functional or biological studies.

Section \ref{BAYESIANINFERENCE} points out the quantities that are of
primary interest and indicates the way to introduce the priors and the
joint density used for posterior inference; the hybrid MCMC for
sampling the posterior density is too complicated and is postponed to
the supplementary material [\citet{C3}]. Section \ref
{SIMULATIONSTUDIES} presents simulation studies to demonstrate that
models having shRNA specific $t$-distributions outperform models assuming
constant variance Gaussian error terms and to explore the extra-power a
RNAi HTS with replicates has in identifying shRNAs affecting pathway
activity and in estimating the false discovery rates. Section \ref
{DATAPREPROCESSING} summarizes the data preprocessing procedures for
data from RNAi HTS; these include edge effect adjustment,
normalization, and outlier removal. Section \ref{DATAANALYSIS}
analyzes the data from the HCV study. In addition to illustrating the
methods in real data analysis, evaluating the performance of the
methods by negative control wells, and proposing shRNA lists with
associated false discovery rates, we compare our results with those
from a limited Q-PCR study and those using the standard $Z$-score method;
we also compare our results with those from similar RNAi HTS in the
literature. The former indicates that results based on our methods are
in better agreement with those based on Q-PCR than those based on
$Z$-score are and the latter spotted some common findings, which is
encouraging in view of the lack of concordance in this area pointed out
in \citet{C4}. Section \ref{DISCUSSION} gives a brief discussion of
future investigations.

\section{Bayesian inference}\label{BAYESIANINFERENCE}

With the likelihood in (\ref{eqlikelihood}), we now describe
a~hierarchical model for Bayesian inference. The following quantities are
of primary interest. The posterior probability that $\gamma_i=0$ given
the data, denoting $p_i=P(\gamma_i=0|x,y)$, reports the probability of
incurring no activity change by the $i$th shRNA; smaller $p_i$ suggests
larger probability of incurring activity change; $1-p_i$ gives the
probability of incurring activity change. The marginal posterior
distribution of $\beta_i$ indicates the amount of positive or negative
influence on the pathway activity given that there is activity change,
which is denoted by $\pi(\beta_i |\gamma_i=1,x,y)$. We will see in our
real data analysis that $E(\beta_i |\gamma_i=1,x,y)$ is highly
correlated with $p_i=P(\gamma_i=0|x,y)$ and is often very useful in
providing a list for further study.

The priors on ($\gamma$, $\beta$), $\mu$, $\alpha_0$, $\alpha_1$,
$\sigma_x^2$, $\sigma_y^2$, $d_x$, and $d_y$ are independently
assig\-ned. Let $N(\cdot|0,V)$ denote the normal density with mean zero
and~varian\-ce~$V$. Let $\operatorname{IG}(\cdot|A,B)$ denote the inverse gamma density
with shape\vspace*{1pt} parame\-ter~$A$ and scale parameter $B$. The prior on $\sigma
_x^2$ is defined by assuming~$\sigma_{x_1}^2,\ldots,\sigma_{x_I}^2$ an
i.i.d. sequence with density $\operatorname{IG}(\sigma_{x_i}^2|A_x,B_x)$ and that on
$\sigma_y^2$ is defined similarly by assuming $\sigma_{y_1}^2,\ldots,\sigma
_{y_I}^2$ an i.i.d. sequence with density $\operatorname{IG}(\sigma_{y_i}^2|A_y,B_y)$.
Let $A_x=a_x^2 / b_x +2$, $B_x=(a_x^2 / b_x +1)a_x$, $A_y=a_y^2 / b_y
+2$, and $B_y=(a_y^2 / b_y +1)a_y$, then $\sigma_{x_i}^2$ has mean
$a_x$ and variance $b_x$ and $\sigma_{y_i}^2$ has\vspace*{1pt} mean~$a_y$ and
variance~$b_y$. The prior is assigned by assuming $a_x$, $b_x$,~$a_y$,
and~$b_y$ are independent with uniform distribution $\mathcal{U}(0,\phi
_1)$, $\mathcal{U}(0,\phi_2)$, $\mathcal{U}(0,\phi_3)$, and~$\mathcal
{U}(0,\phi_4)$, respectively. We note that while the model allows shRNA
specific variance, information is shared between them through these
distributions so as to stabilize the variances. We will make use of the
fact that we have replicates in the experimental design to choose $\phi
_1$ and $\phi_2$. Roughly speaking, we choose~$\phi_1$~($\phi_2$) equal
to or larger than twice the sample mean (sample variance) of $\{
(x_{i1}-x_{i2})^2/2 | i=1,\ldots,I\}$, because we wish to have a more
vague prior. The way to choose $\phi_3$ and $\phi_4$ is similar. The
prior for~$d_x$ and~$d_y$ are uniform on the integers from $1$ to $100$.

The prior on $\gamma$ is defined by assuming $\gamma_1$, $\gamma
_2,\ldots, \gamma_I$ an i.i.d. sequence with $P(\gamma_i=0)=1-P(\gamma
_i=1)=p$, which is assumed to have a density~$\operatorname{\mathcal{B}eta}(\phi_5,\phi
_6)$. If $\gamma_i=0$, let $\beta_i=0$; otherwise, let $\beta_i$ have a
density $N(\beta_i|0,V)$. Both $\alpha_0$ and~$\alpha_1$ also have
prior densities $N(\alpha_0|0,V)$ and $N(\alpha_1|0,V)$. We assume $V$
has inverse gamma density $\operatorname{IG}(V|\phi_7,\phi_8)$. The parameters $(\phi
_5,\phi_6)$ studied are~$(9,1)$ and $(1,1)$; we will see in Section
\ref{Sensitivityanalysis} that the posterior inferences do not seem to
be sensitive to the values of~$(\phi_5,\phi_6)$. We choose~$\phi_7$ and~$\phi_8$
to make $\operatorname{IG}(V|\phi_7,\phi_8)$ have large mean and variance so
that it is less informative. We note that the prior on $(\gamma_i, \beta
_i)$ follows and extends that in \citet{S1} for microarray
gene expression studies.

The prior for $\mu$ is defined by assuming $\mu_1$, $\mu_2,\ldots,
\mu _I$ an i.i.d. sequence with uniform distribution
$\mathcal{U}(-3,1)$, which is motivated by the fact that in the HCV
study, $\log(\tilde {x}_{ij})$ belongs to $(-3,1)$ for every $i$ and
$j$; see Figure~\ref{RawAndLogDataPlot}(b). Section
\ref{Sensitivityanalysis} also shows that our main results do not vary
much with the changes in the prior on~$\mu$.

In addition to the above rationale, we limit our attention to the
priors for which the models fit the data well by posterior predictive
check, discussed by \citet{G2}. In case there are several
models fitting the data well, we prefer the less informative priors.

It follows from (\ref{eqlikelihood}) in Section \ref{Alog-linear}
that the joint density of ($x$, $y$, $\omega_x$, $\omega_y$, $\gamma$,
$\beta$, $\mu$, $\alpha_0$, $\alpha_1$, $\sigma_x^2$, $\sigma_y^2$,
$d_x$, $d_y$, $p$, $V$, $a_x$, $b_x$, $a_y$, $b_y$) is
%
\begin{eqnarray}\label{eqposterior}
&& g(x, y, \omega_x, \omega_y| \gamma, \beta, \mu, \alpha_0, \alpha_1,
\sigma_x^2, \sigma_y^2, d_x, d_y) \nonumber\\
&&\qquad{}\times p^{I-\sum_{i=1}^I \gamma_i +
\phi_5 -1}(1-p)^{\sum_{i=1}^I \gamma_i + \phi_6 -1}
\nonumber\\
&&\qquad{}\times\prod_{i=1}^I (N(\beta_i|0,V))^{\gamma_i} \cdot N(\alpha
_0|0,V) \cdot N(\alpha_1|0,V) \cdot \operatorname{IG}(V|\phi_7,
\phi_8) \\
&&\qquad{}\times\prod_{i=1}^I \operatorname{IG}(\sigma_{x_i}^2|A_x,B_x) J(A_x,B_x) \nonumber\\
&&\qquad{}\times\prod
_{i=1}^I \operatorname{IG}(\sigma_{y_i}^2|A_y,B_y)
J(A_y,B_y),\nonumber
\end{eqnarray}
where $J(\cdot,\cdot)$ is the Jacobian matrix of the transformation
from $(A,B)=(a^2 / b +2,(a^2 / b +1)a)$ to $(a,b)$. It is based on (\ref
{eqposterior}) that we propose a hybrid MCMC algorithm for computing
the posterior distribution [see, e.g., \citet{R1}, page 393]. We note that we use the trick to update $\gamma_i$
after integrating out $\beta_i$ to make the algorithm more efficient;
see \citet{G3}. Several useful observations that
accelerate the hybrid MCMC algorithm and the algorithm itself are given
in the supplementary material [\citet{C3}] to streamline the presentation.

\section{Simulation studies}\label{SIMULATIONSTUDIES}

The purposes of these simulation studies are to evaluate the
performance of our Bayesian method and to indicate the benefits of
having replicates in a RNAi HTS experiment. Our studies show that the
hierarchical model allowing error terms having $t$-distributions and
shRNA specific variance outperform the usual Gaussian model with fixed
common variance in terms of study power and false positive rate
estimation.

The first data set we study is generated as follows. The total number
of shRNAs is $6\mbox{,}130$. The number of replicates is $2$. The model index
parameter~$\gamma_i$ is equal to~$1$ for $i=1,\ldots,100$, and equal to
$0$ for $i>100$. If $\gamma_i=0$, then $\beta_i=0$; otherwise generate
$\beta_i$ from uniform distribution on $(-5,3)$. Let the parameters $\mu
_i$ be generated by $(0.248+2.77) \cdot\operatorname{\mathcal{B}eta}(6,2)-2.77$,
which has support on $(-2.77,0.248)$ and is very close to the empirical
distribution of~$\{ x_{ij} \}$ in the HCV study.\vspace*{1pt} Let $\alpha_0=12.557$,
$\alpha_1=2.538$, $d_x=3$, and $d_y=3$. Let $\sigma_{x_i}^2$ be
$\operatorname{IG}(\sigma_{x_i}^2|3,0.2)$, which has mean $0.1$ and variance $0.01$,
and $\sigma_{y_i}^2$ be $\operatorname{IG}(\sigma_{y_i}^2|3,1)$, which has mean $0.5$
and variance $0.25$; we note that these means and variances are many
times larger than those suggested by the data in the HCV study, given
in Section \ref{Analysisandvalidation}. In fact, some aspects of
the data set are chosen deliberately so they are similar to those in
the HCV study and some are chosen to be distinguished from those in the
model or prior specifications in the previous sections. The aspects
similar to the HCV study include the number of shRNAs, the number of
replicates, and the distribution of the cell viabilities. These help to
make the simulation studies relevant to real data analysis and are
useful for the study of robustness of our method.

\begin{figure}
\begin{tabular}{@{}cc@{}}

\includegraphics{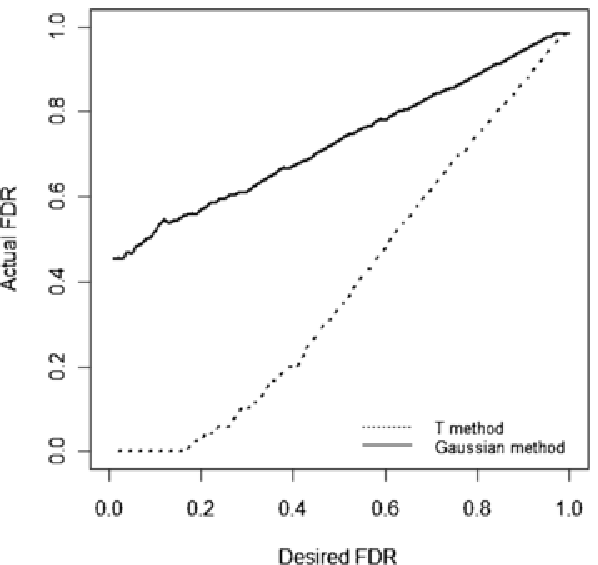}
  & \includegraphics{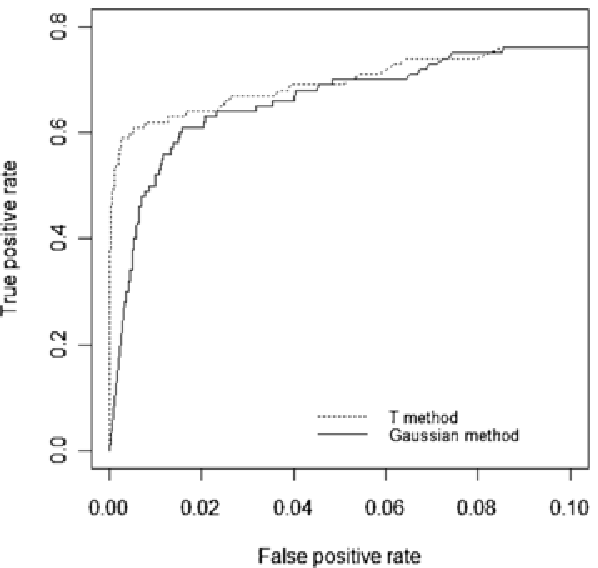}
\\
(a) & (b)
\end{tabular}
\caption{Comparison of the results analyzed by the T method and the
Gaussian method, based on data generated from $t$-distributions with
$J=2$. \textup{(a)} The actual FDR versus the desired
FDR; \textup{(b)} the ROC curves.} \label{SimPlot1}
\vspace*{-6pt}
\end{figure}

Our first analysis uses the model and prior described earlier in
Sections~\ref{Alog-linear} and~\ref{BAYESIANINFERENCE}; in particular,
we choose $\phi_1=\phi_2=\phi_4=0.2$, $\phi_3=\phi_6=1$, $\phi_5=9$,
\mbox{$\phi_7=3$}, and $\phi_8=30\mbox{,}000$. For the second analysis, we use
the model in Section~\ref{Alog-linear} with $\omega
_{x_{ij}}=\omega_{y_{ij}}=1$, $\sigma_{x_i}^2=\sigma_x^2$, and $\sigma
_{y_i}^2=\sigma_y^2$, both of which are estimated from the data, and a
constant $V$, which is equal to the posterior mean of the $V$ obtained
in the first analysis. We summarize the comparison of these two
analyses in Figure~\ref{SimPlot1}, in which the first analysis is
called the~T method and the second is termed the Gaussian method.
Figure \ref{SimPlot1}(a) provides plots of actual false discovery rates
against the desired false discovery rates, which was used in Lo and
Gottardo (\citeyear{L3}). Figure~\ref{SimPlot1}(b) shows plots of the
true positive rates against the false positive rates. It is clear from
Figures~\ref{SimPlot1}(a) and~(b) that the method allowing
$t$-distribution and shRNA specific variance performs much better.

%

\begin{figure}
\begin{tabular}{@{}cc@{}}

\includegraphics{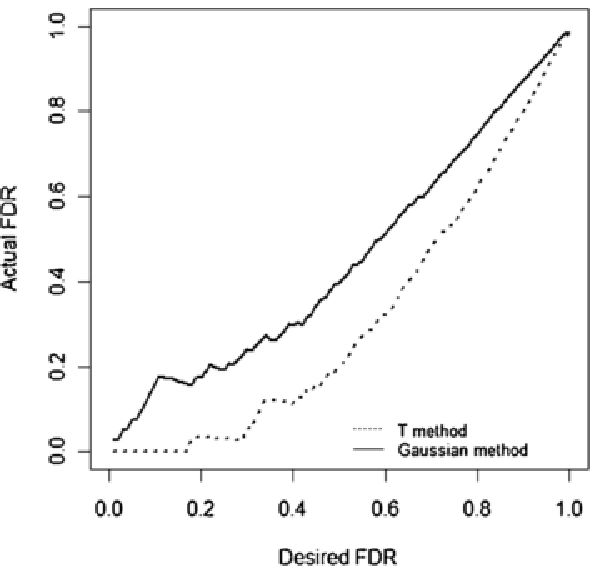}
  & \includegraphics{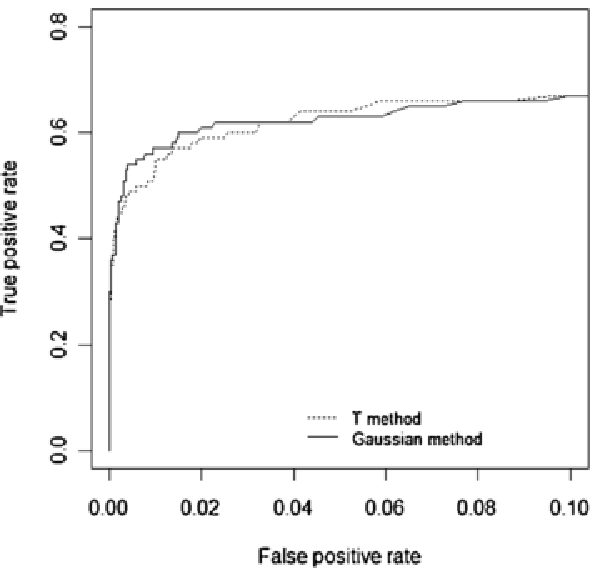}
\\
(a) & (b)
\end{tabular}
\caption{Comparison of the results analyzed by the T method and the
Gaussian method, based on data generated by normal distribution with
$J=2$. \textup{(a)} The actual FDR versus the desired
FDR; \textup{(b)}~the ROC curves.} \label{SimPlot2}
\vspace*{-3pt}
\end{figure}

The second data set is generated by the same model parameters as the
first data set except $\omega_{x_{ij}}=\omega_{y_{ij}}=1$, $\sigma
_{x_i}^2=0.1$, and $\sigma_{y_i}^2=0.5$; thus,\vspace*{1pt} the error
terms are normal with the same variance. This data set is again
analyzed by the above-mentioned T method and the Gaussian method and a
comparison of the two is summarized in Figure \ref{SimPlot2}. We can
see that the Gaussian method performs only slightly better.

%

\begin{figure}
\begin{tabular}{@{}cc@{}}

\includegraphics{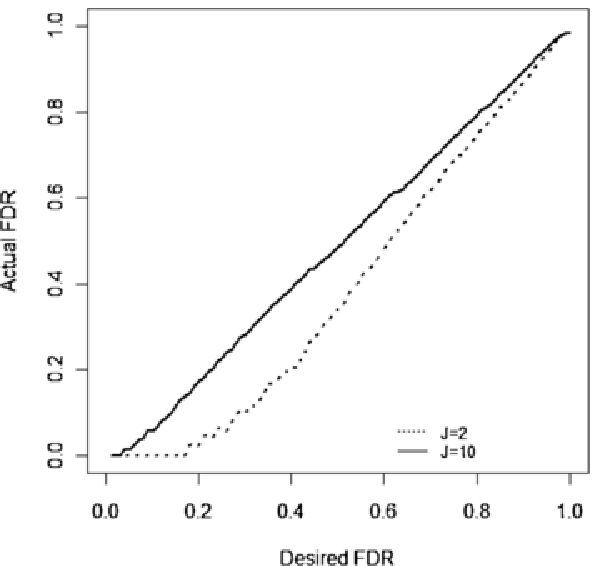}
  & \includegraphics{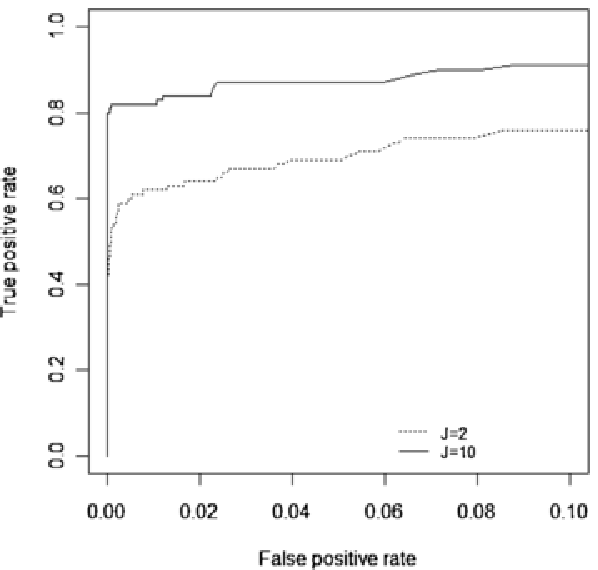}
\\
(a) & (b)
\end{tabular}
\caption{Comparison of the results with $J=2$ and $J=10$, based on data
generated from $t$-distributions and analyzed by the T method.
\textup{(a)} The actual FDR
versus the desired FDR; \textup{(b)}
the ROC curves.} \label{SimPlot3}
\end{figure}

The third data set is generated in the same way as the first except
that the number of replicates $J=10$. Figure \ref{SimPlot3} compares
the results obtained from the data with $J=2$ and $J=10$, both of which
are analyzed by the T method of this paper. Figure \ref{SimPlot3} shows
that the results using $J=10$ is much better than $J=2$.%

%

In fact, we also compared these two methods using data generated with
error terms being neither Gaussian nor $t$-distributed and observed
results similar to those in Figures \ref{SimPlot1} and \ref{SimPlot3}.
To keep the paper concise, we omit reporting these
comparisons.

\section{Data preprocessing}\label{DATAPREPROCESSING}

Data preprocessing for RNAi HTS usually consists of three parts: edge
effect adjustment, normalization to reduce plate effects, and outlier
removal. Several data preprocessing methods have been proposed to deal
with these issues; see, for example, \citet{B4},
\citet{M1} and \citet{Z2}. These methods are adapted
to suit the present experiment.

The following three observations motivate our data preprocessing
procedure. The first observation is that technically replicated
measurements should be close to each other; this provides an
opportunity to evaluate the general quality of the experiment as well
as to eliminate shRNAs showing large discrepancy between the replicated
measurements. The second observation is that the assignment of a shRNA
to a well and the effect of this shRNA on the measurements are
independent; our edge-effect adjustment makes use of this observation.
The third observation is that measurements of control wells of the same
type are expected to be close to each other; the measurements for NTNP
wells are expected to be larger than those for SN wells and those for
NTWP wells are expected to have the smallest measurements; violation of
these indicates poor quality of the experiments. In the HCV study, no
such violation is observed in any single plate either for luciferase
activity or cell viability; our normalization procedure makes use of
this observation.

Our data preprocessing are performed by first conducting edge effect
adjustment and selecting the control wells so as to perform
normalization, and then deleting the outliers. The final data in the
analysis consists of $6\mbox{,}130$ shRNA and $419$ negative control wells,
each of which has four replicates.

Control wells serve two purposes in our data preprocessing. Since the
measurements for NTNP wells are larger than those for SN wells and NTWP
wells have the smallest measurements in any plate, it seems a good idea
to use these control wells to perform normalization if we can first
delete outliers in the control wells. This outlier deletion is carried
out for luciferase activity and cell viability separately. Here we
present the deletion criterion for the luciferase activity in a shRNA
or control well; that for cell viability is similar and omitted. Using
the fact that there are four platewise replicates with two of them
measuring luciferase activity, we consider the ratio of the difference
of its two replicated measurements to its mean for a given well
position, and delete both of its luciferase activities for a given well
position if the ratio is large. For the HCV study data, two NTNP
control wells were deleted based on the cell viability measurement
before normalization.

Edge effect exists in both the measurements of luciferase activity and
cell viability; the same adjustment procedure is carried out separately
for each of these two measurements and described as follows. We
partition all the wells in this experiment into three groups: G1
consists of all the wells on the boundary of a plate; G2~consists of
all the wells not in G1 but immediately adjacent to wells in G1; G3~the
remaining wells. A fixed constant is added to the measurement from each
well in G2 so that the average of all the measurements from wells in G2
is equal to that from wells in G3; a similar procedure is applied to
the measurements from each well in G1, except those from NTWP control
wells. The rationale for not performing edge effect adjustment to the
NTWP control wells is that the total numbers of cells in these wells
are very small already and we expect little edge effects there. We note
that all the control wells are in G1.

After the outliers in the controls have been removed and the edge
effect has been taken into account, we perform normalization to reduce
the plate effect. The same normalization procedure is done separately
for luciferase activity and cell viability; we describe the procedure
for luciferase activity and omit that for cell viability, because of
the similarity. Our normalization will, in particular, make mean
measurements of the controls of the same type in each plate taking the
same value for different plates. The procedure is as follows. We denote
the mean measurements of NTNP wells, SN wells, and NTWP wells in plate
$h$ by $a_h$, $b_h$, and $c_h$, respectively. Let~$a$,~$b$, and~$c$
denote, respectively, the mean measurements of all of the NTNP wells,
SN wells, and NTWP wells from all the plates in this experiment. The
normalization procedure for wells in plate $h$ is accomplished by using
the unique continuous piecewise\vadjust{\goodbreak} linear function that maps $a_h$, $b_h$,
and $c_h$ to~$a$,~$b$, and~$c$, respectively. Namely, its value at $x$ is
\[
\cases{x-c_h+c, &\quad if $x\leq c_h$; \cr
(a-c)(x-c_h)/(a_h-c_h)+c, &\quad if $c_h < x \leq a_h$;\cr
(b-a)(x-a_h)/(b_h-a_h)+a, &\quad if $a_h < x \leq b_h$;\cr
x-b_h+b, &\quad if $b_h < x$.}
\]
The luciferase activity of a well in plate $h$ is to be replaced by the
value of the piecewise linear function at that luciferase activity. In
the plates without NTWP wells, the normalization is carried out using
the piecewise linear function defined by $a$ and $b$.

After normalization, we exclude a shRNA from the analysis if either the
ratio of the difference of its two replicated luciferase activity to
its mean or the ratio corresponding to its two replicated cell
viability measurements is large. For each of the two measurements, we
excluded two percent of the shRNAs that have the most extreme ratios in
the HCV study.

\section{The HCV study}\label{DATAANALYSIS}

We now apply our methods to analyze the HCV study data. In particular,
we will provide lists of candidate shRNAs causing the reduction of HCV
replication without changing cell viability. We will first explain in
some detail the way we apply the methods in analyzing the data and the
way we use part of the control wells to evaluate the performance of the
methods. We will also examine our methods in light of a limited Q-PCR
experiment, compare with the results using the standard $Z$-score
approach, show that our Bayesian analysis is insensitive to the choice
of prior, and indicate that there does exist concordance between our
findings and those in \citet{S2} and \citet{T1}.

\subsection{Data analysis}\label{Analysisandvalidation}

After the data preprocessing procedures, we analyzed the data using the
Bayesian method of this paper. Following the data analysis strategies
described in Section \ref{BAYESIANINFERENCE}, we choose $\phi_1=\phi
_2=0.03$ and $\phi_3=\phi_4=0.2$. In fact, $\{ (x_{i1}-x_{i2})^2/2 |
i=1,\ldots,6\mbox{,}549\}$ has mean $0.0016$ and variance $8\times10^{-6}$ and
$\{ (y_{i1}-y_{i2})^2/2 | i=1,\ldots,6\mbox{,}549\}$ has mean $0.0107$ and
variance $0.0004$. Except for the sensitivity analysis in Section
\ref{Sensitivityanalysis}, we always use $\phi_5=9$, $\phi_6=1$, $\phi
_7=3$, and $\phi_8=1.9\times10^5$ in this section. We note that
$\operatorname{IG}(\cdot|3,1.9\times10^5)$ has mean $95\mbox{,}000$ and variance
$9.025\times10^9$.

We use about half of the SN and NTNP wells and all of the NTWP wells
for normalization purposes and use the remaining control wells to
evaluate our analysis method and to help decide the knockdown effects
of specific shRNAs. In each plate, there is about one NTWP well, two SN
wells, and three NTNP wells. We use all the NTWP wells for
normalization. We also note that there are in total $71$ plates and
$426$ control wells in one replicate and there are four platewise
replicates, as described at the end of Section \ref
{RNAinterference}. After the initial outlier deletion of $2$ NTNP
controls, we use $277$ controls for normalization.\vadjust{\goodbreak} The outlier deletion
step after normalization excluded another $6$ controls, three of which
were among those used for normalization earlier. These outliers are
excluded from the analysis. It turns out that our analysis includes
$419$ controls, $274$ of which were used for normalization and the
remaining $145$ controls are used to evaluate our methods as well as to
decide the knockdown effect of specific shRNAs.

The posterior distributions are obtained by the hybrid MCMC algorithm
in the supplementary material [\citet{C3}]. We calculated the
Gelman--Rubin statistic $\hat{R}$ for all the important estimands based
on five chains with random initial values in several studies and found
all the $\hat{R}$'s were less then $1.1$ using the iterations between
$10\mbox{,}000$ and $20\mbox{,}000$. Based on this experience and the fact that all
the studies in this section involve similar computations, we run one
chain with $40\mbox{,}000$ iterations and use the latter $20\mbox{,}000$ iterations
to calculate the posterior distributions of all the parameters in each study.

%
\begin{figure}
\includegraphics{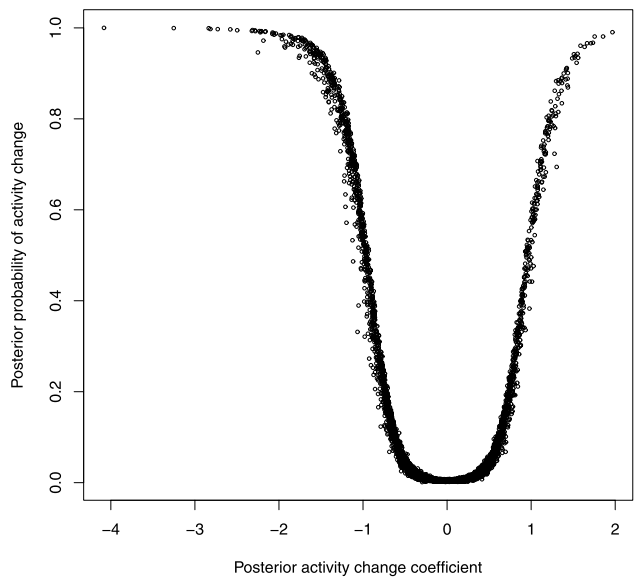}%

\caption{The volcano plot of posterior activity change coefficient and
posterior probability of activity change for the HCV study.}
\label{pVSbetaRealData}
\end{figure}

Figure \ref{pVSbetaRealData} gives some idea of the distribution of
the activity change in this experiment. Each dot in Figure \ref
{pVSbetaRealData} gives the results of one shRNA; its vertical
coordinate gives its posterior probability of incurring activity change
$1-p_i$ and its horizontal coordinate gives its magnitude of activity
change $E(\beta_i | \gamma_i=1,x,y)$, given there is activity change.
$E(\beta_i | \gamma_i=1,x,y)$ is also called the posterior activity
change coefficient. Figure \ref{pVSbetaRealData} indicates a very
clear relation between these two quantities and it will be clear that
both are useful in examining the activity change.

%
\begin{figure}

\includegraphics{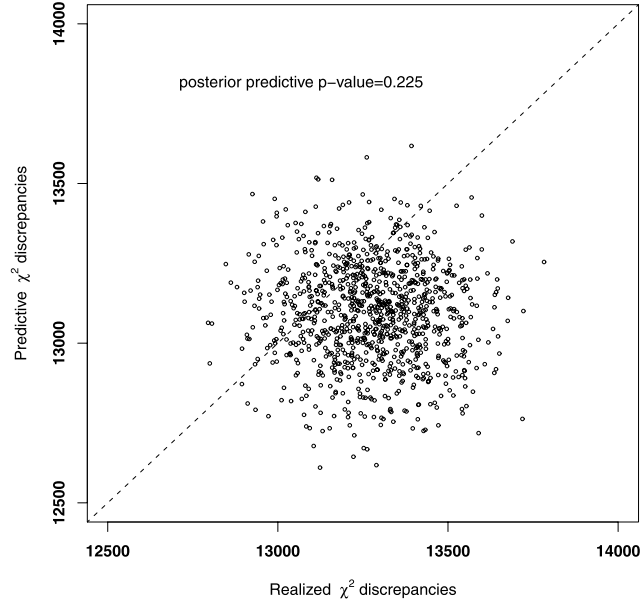}

\caption{The Scatterplot of predictive versus realized $\chi^2$
discrepancies for the HCV study.} \label{Scatterplot}
\end{figure}

Figure \ref{Scatterplot} gives\vspace*{1pt} the graphical display of the posterior
predictive check that using the $\chi^2$ omnibus discrepancy test quantity,
\[
\sum_{i=1}^{6\mbox{,}549} \sum_{j=1}^2 \frac{(y_{ij} - \alpha_0 - \gamma_i \beta
_i - \alpha_1x_{ij})^2}{\alpha_1^2 \sigma_{x_i}^2d_x/(d_x-2) + \sigma
_{y_i}^2d_y/(d_y-2)}.
\]
The concept of posterior predictive check appeared in \citet{G2}; see also \citet{G1}.
Figure \ref{Scatterplot} seems
to indicate that the model fits the data quite well. In fact, in the
choice of priors, we find large $\phi_8$ not only makes $V$ less
informative but also make model fit possible. By the way, the posterior
means of $\alpha_0$, $\alpha_1$, $V$, $d_x$, and $d_y$ are,
respectively, $12.557$, $2.538$, $58.207$, $30$, and~$10$.

We may regard the minimal interval containing the posterior means of
the cell viability $E(\mu_i|x,y)$ of the $274$ controls as the range of
normal cell viability, which is $(-0.635,-0.007)$, although this
definition of normal cell viability might be a little too stringent.
Nevertheless, we find all of the remaining $145$ controls are in this
normal range. This finding seems to suggest that our methods give good
estimates of the cell viabilities. Among all the $6\mbox{,}130+419=6\mbox{,}549$ wells,
$5\mbox{,}862$ have their cell viabilities in the normal range, $602$ of them
less than $-0.635$ and $85$ of them larger then $-0.007$.

\begin{figure}

\includegraphics{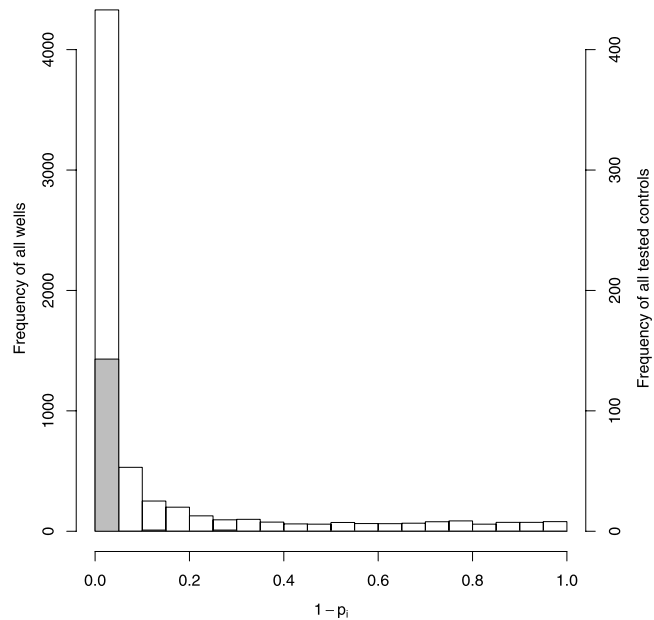}

\caption{Histogram of posterior probability of activity change $1-p_i$.
The dark histogram, referred to as Histogram 7\textup{(a)}, is for the $145$
control wells reserved for evaluation purposes. The light histogram,
referred to as Histogram 7\textup{(b)}, is for all the $6\mbox{,}549$ wells.}
\label{Histpi}
\end{figure}

Figure \ref{Histpi} reports two histograms of the probability of
activity change $1-p_i$. The dark one, called Histogram 7(a), is the
histogram for the $145$ controls; the light\vadjust{\goodbreak} one, called Histogram 7(b),
is that for all the $6\mbox{,}549$ wells. Since Histogram~7(a)
concentrates heavily on the left, we think our methods assigned
appropriate posterior probability to these $145$ control wells.

\begin{figure}

\includegraphics{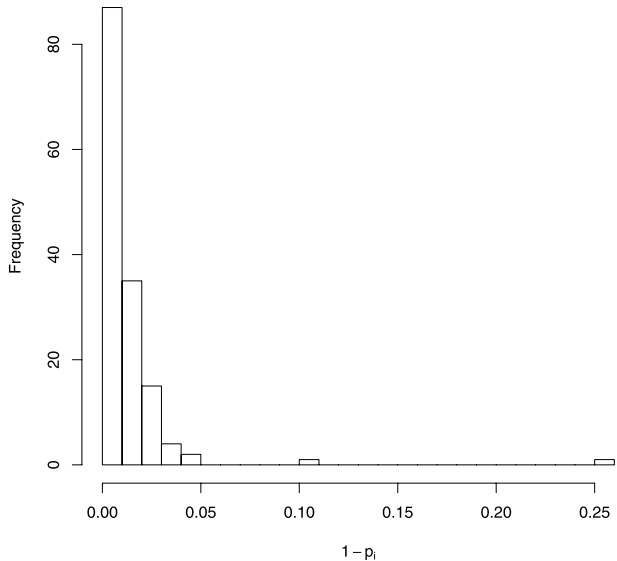}

\caption{Histogram of the posterior probability of activity change
$1-p_i$ for the $145$ control wells.} \label{HistpiRegularViability}
\end{figure}

In fact, Histogram 7(a) provides opportunities for the study of false
discovery rate. For this, we present the data of Histogram 7(a) in Figure
\ref{HistpiRegularViability} using finer resolution so as to exhibit
more details. Figure \ref{HistpiRegularViability} seems to suggest that
the set of shRNAs having $1-p_i$ larger than $0.2532$, $0.1055$, or
$0.05$ has false discovery rate, respectively, $0$, $1/145$, or
$2/145$, among others.

%

%

\begin{figure}

\includegraphics{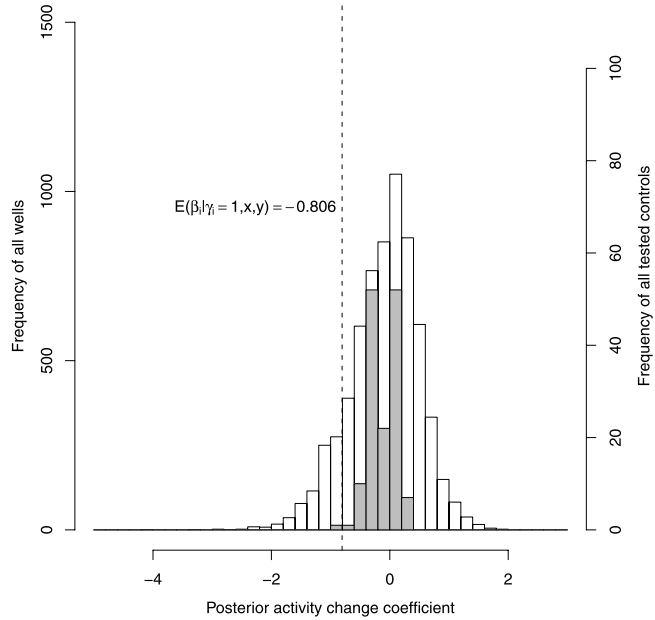}%
\vspace*{-3pt}
\caption{Histogram of posterior activity change coefficients. The dark
histogram is for the $145$ control wells. The light histogram is for
all the $6\mbox{,}549$ wells.} \label{HistpiActivityChange}
\vspace*{-3pt}
\end{figure}

There is still another quantity that is useful in assessing the
performance of our methods, namely, the posterior activity change
coefficient $E(\beta_i | \gamma_i=1,x,y)$. Figure \ref
{HistpiActivityChange} provides two histograms of posterior activity
change coefficient; the dark one, sitting above the interval $(-0.806,
0.392)$, is that for the $145$ shRNAs in the control wells and the
light one, sitting above the interval $(-4.081, 1.967)$, is that for
the total set of $6\mbox{,}549$ shRNAs. The fact that the former sits in the
middle of the latter, as expected, seems to give another piece of
indication that the experiment and data analysis are reliable.

\subsection{Main results}\label{Mainresults}
We are now in a position to provide shRNA lists,~based on which the
laboratory scientists can conduct experiments to identify those that
affect HCV replication without affecting cell viability. Knowing
that
among the $785$ shRNAs having their $E(\beta_i | \gamma_i=1,x,y)$ less
than $-0.806$, $601$ of them have their cell viability in the normal
range $(-0.635,-0.007)$, we will start the selection with this set of
$601$ shRNAs. For this, we will~make~use of the direct posterior
probability approach discussed in \citet{N1}.

%

Assuming that placing a shRNA having posterior probability of not
incurring activity change $p_i$ in the list for further studies creates
a loss of the amount $p_i$, we rank the shRNAs according to their $p_i$
and form the list as follows. Given a set $\varsigma$ of shRNAs for
which the losses are less than a positive number $\kappa$, we define
\[
C(\kappa)=\sum_i p_i 1_{[p_i<\kappa]},
\]
which is the sum of the losses for all the shRNAs in $\varsigma$. We
note that $C(\kappa)/|\varsigma|$ is called the posterior false
detections rate (PFDR), where $|\varsigma|$ is the size of the
set
$\varsigma$; we also note that with the shRNAs ordered by $p_i$, there
is a~natural correspondence between $\kappa$ and set size $|\varsigma|$
and that $C(\kappa)/|\varsigma|$ increases with~$\kappa$. To obtain a
list of shRNAs having PFDR $C(\kappa)/|\varsigma|$ less than $\alpha$,
we can find the largest possible $\kappa$ or $|\varsigma|$ such that
$C(\kappa)/|\varsigma|$ is less than $\alpha$. In practice, we can
either fix the list size and then compute its PFDR or fix the PFDR and
compute the list size. For the set of $601$ shRNAs having normal cell
viability and posterior activity change coefficient less than $-0.806$,
Table~\ref{PFDRForKnownDown} is a summary of various lists of shRNAs.
Its left column reports PFDR by first fixing\vadjust{\goodbreak} the set size and its right
column reports list size by first fixing the PFDR. For example, the
second row in the left column says that the list of top $100$ candidate
shRNAs, targeting $77$ genes, has PFDR $0.047$; the second row of the
right column says given PFDR being $0.05$, the list has $104$ shRNAs,
targeting $81$ genes, and one of the genes has three of the $104$
shRNAs targeting it, $21$ of the genes have two such shRNAs targeting
each of them and the remaining $59$ genes have only one such shRNA
targeting each of them.

%
\begin{table}[b]
\caption{Posterior false detection rates for lists of shRNAs whose
$E(\mu_i|x,y) \in(-0.635,-0.007)$ and $E(\beta_i | \gamma
_i=1,x,y)<-0.806$ in the HCV study}\label{PFDRForKnownDown}
\begin{tabular*}{\tablewidth}{@{\extracolsep{\fill}}lcccrrrrr@{}}
\hline
\multicolumn{2}{@{}c}{\textbf{Fix list size}}
& \multicolumn{7}{c@{}}{\textbf{Fix PFDR}} \\
\ccline{1-2,3-9}\\[-8pt]
&\multirow{2}{60pt}{\centering{\textbf{Number of genes~(shRNAs)}}}
& & \multirow{2}{60pt}{\centering{\textbf{Number of genes~(shRNAs)}}}
&\multicolumn{5}{c}{\textbf{Number of shRNAs}}\\
\cline{5-9}\\[-8pt]
\textbf{PFDR} & & \multicolumn{1}{c}{\textbf{PFDR}}
& &\multicolumn{1}{c}{\textbf{5}}&\multicolumn{1}{c}{\textbf{4}}&\multicolumn{1}{c}{\textbf{3}}
&\multicolumn{1}{c}{\textbf{2}}&\multicolumn{1}{c@{}}{\textbf{1}}\\
\hline
0.024& \hphantom{0}40 (50)\hphantom{0} & 0.01& 20 (21) &&&&1&19\\
0.047 & \hphantom{0}77 (100) & 0.05& \hphantom{0}81 (104) &&&1&21&59\\
0.106 &151 (200) & 0.10&144 (190) &&2&6&27&109\\
0.174 &210 (300) & 0.20&234 (337) &1&3&15&57&158\\
0.242 &271 (400) & 0.30&309 (488) &3&7&31&79&189\\
\hline
\end{tabular*}
\end{table}

We note that when we are interested in individual shRNAs, those in the
first row are better candidates than those in the other rows; when we
are interested in genes that affect HCV replication, then we need to
consider the number and the knockdown effects of all the shRNAs in each
gene. We will illustrate the latter in Section \ref{Somecomparison}.

\citet{T1} pointed out the issue of the choice of threshold in
the $Z$-score approach to hit selection; higher threshold causes little
overlap with other studies and lower threshold results in too many
siRNAs for secondary validation. Our approach benefits from not only
the statistical model but also the negative controls. The above
analysis points out an issue regarding the number of negative controls
to be used for the false discovery rate estimate. We believe that if we
had more negative controls, we could have provided more information,
including false discovery rate, for the selection of shRNAs or genes.\vadjust{\goodbreak}

For the purpose of comparison in Sections \ref{Comparisonwith} and
\ref{Somecomparison}, we say a shRNA in the HCV study causes activity
change if either its $p_i<1-0.2532=0.7468$ or its $E(\beta_i | \gamma
_i=1,x,y) \notin[-0.806,0.392]$; a shRNA has normal cell viability if
its $E(\mu_i|x,y) \in(-0.635,-0.007)$.

\subsection{Q-PCR validation}\label{Q-PCR}
Q-PCR is one of the popular methods to study activity change in a
secondary analysis of a gene list from a primary RNAi HTS. In this
study, both HCV RNA and house-keeping gene beta-actin are reverse
transcribed and then quantified using Q-PCR in a well with a given
shRNA or without any shRNA transduction. The ratio of the quantity of
HCV RNA in a well with the shRNA transduction to that in a well without
any shRNA transduction and the corresponding ratio for beta-actin are
first calculated; the former ratio divided by the latter ratio gives
the activity change measured by Q-PCR for shRNA $i$, which is denoted
by $k_i$.

When $k_i$ is around $1$, we may say there is no activity change; when
it is large (small), we may say it increases (decreases) the activity.
Although Q-PCR is considered to provide reliable measurements, we do
not know of a good threshold on $k_i$ for declaring activity change.
Thus, we choose to present the correlation between the activity $k_i$
reported by the Q-PCR assay and posterior activity change coefficient
reported by our methods as a way to make the comparison.

We examined by the Q-PCR method the pathway activity of $86$ shRNAs in
the HCV study; we note that $68$ of them have their ratio for
beta-actin belonging to the interval $(0.7, 1.43)$; these $68$ shRNAs
may thus be considered to have less appreciable effect on cell
viability. The empirical correlations of these two assays are $0.68$
for all these $86$ shRNAs and $0.77$ for the $68$ shRNAs not affecting
cell viability. We note that these $86$ shRNAs had been chosen for
Q-PCR assay before the methods of this paper were proposed. We note
that we do not expect perfect correlation, because, for HCV
replication, our RNAi\vadjust{\goodbreak} HTS monitors the luciferase protein and Q-PCR
measures HCV RNA and, for cell viability, ours monitors the level of
dehydrogenase and Q-PCR montiors the level of mRNA of beta-actin.

%
\begin{figure}[b]
\begin{tabular}{@{}cc@{}}

\includegraphics{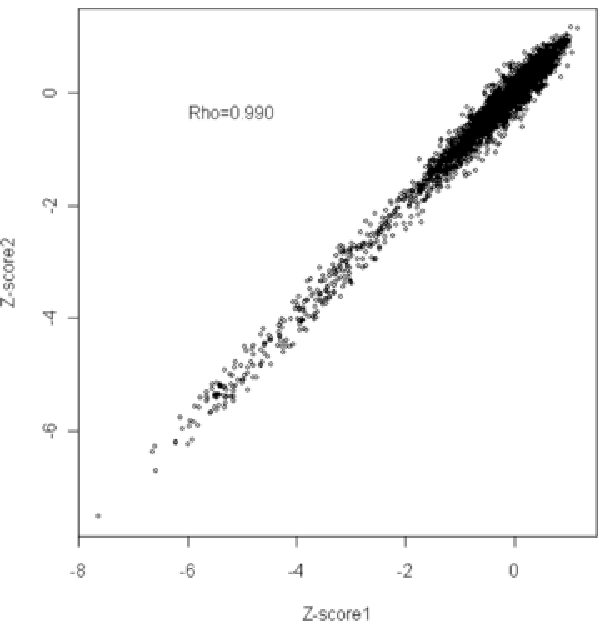}
  & \includegraphics{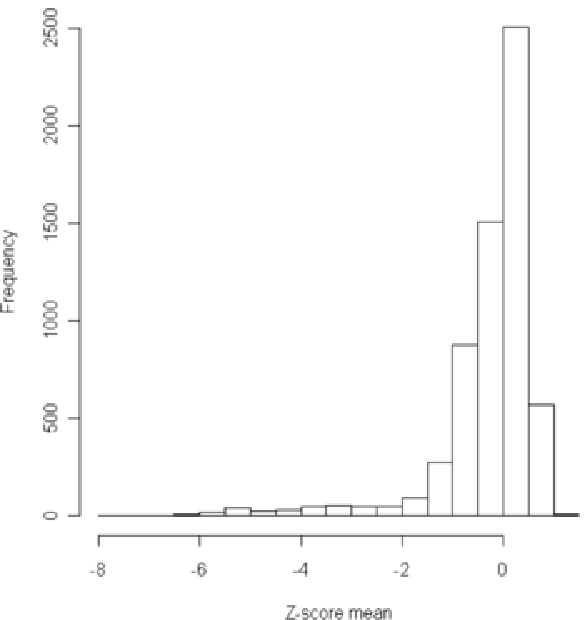}
\\
(a) & (b)
\end{tabular}
\caption{Scatter plot for $Z$-scores and histogram of $Z$-score mean.
\textup{(a)} The scatter plot for
$(Z_{i1},Z_{i2})$; \textup{(b)} the histogram of
$Z_i=(Z_{i1}+Z_{i2})/2$.} \label{HistZ-score}
\end{figure}

\subsection{Comparison with the $Z$-score approach}\label{Comparisonwith}
Platewise $Z$-score is often used in the primary analysis of RNAi HTS
data; see, for example, \citet{T1}. In particular, \citet{T1} says that
``A $Z$-score is the number of standard deviations of the
experimental luciferase activity above the median plate value.'' Because
the plate effect in the data from the HCV study has been reduced by
normalization, it seems appropriate for us to consider an
experiment-wise $Z$-score and compare this $Z$-score approach with the
method of this paper. We define the $Z$-score as
\[
Z_{ij}=\frac{y_{ij}-M_j}{S_j}.
\]
Here $M_j$ and $S_j$ are, respectively, the median and standard deviation
of $\{ y_{ij}|i=1,\ldots, I \}$ for $j=1,2$, after data preprocessing
procedures.

Hit selection based on $Z$-score chooses shRNAs having extreme $Z$-scores.
It is desirable that these $Z$-scores follow a normal distribution; in
this case, we can conveniently assign a $p$-value to
$Z_i=(Z_{i1}+Z_{i2})/2$ and use $p$-value to describe the extremeness of
the selected shRNA. Figures \ref{HistZ-score}(a) and (b) give, respectively, the plot of
$(Z_{i1},Z_{i2})$ and the histogram of $\{ Z_i|i=1,\ldots, I \}$ for the
HCV study. While Figure \ref{HistZ-score}(a) shows that $Z_{i1}$ and $Z_{i2}$ have
correlation $0.990$ and are in good agreement, indicating both the
assay and the data preprocessing procedure are excellent, Figure \ref{HistZ-score}(b)
suggests they are not normally distributed.\vadjust{\goodbreak}

%
\begin{figure}

\includegraphics{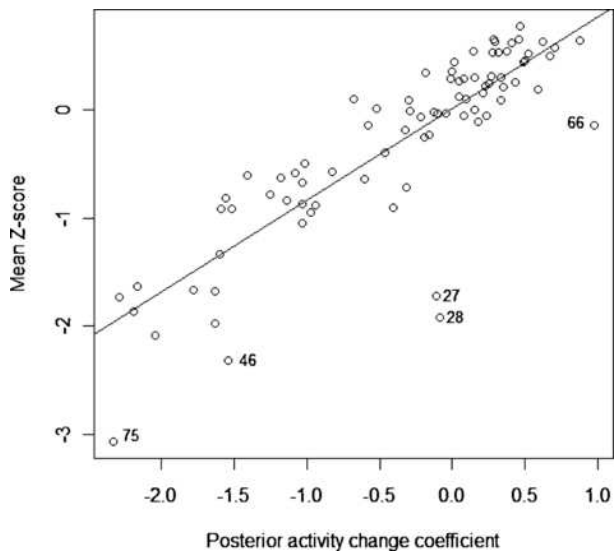}%

\caption{Plot of the $Z$-score vs posterior activity change coefficient
for $86$ shRNAs. The shRNAs that show largest discrepancy between the
$Z$-score and posterior activity change coefficient target, respectively,
\textit{RPS6}, \textit{KIAA1446}, \textit{INPP4A}, \textit{SFRS1},
and \textit{TYK2}. The numerals in the figure
correspond to those in Table \protect\ref{ZscoreLargestD}.}
\label{Z-scorevsBeta}
\end{figure}

For the above set of $86$ shRNAs assayed by Q-PCR and whose subset of
$68$ shRNAs are not affecting cell viability, we find the empirical
correlations of their $Z$-scores $Z_i$ and the activity change measured
by Q-PCR $k_i$ are $0.62$ and $0.72$, respectively. Compared with the
correlations reported in Section~\ref{Q-PCR}, it seems that the
activity change computed by our method is in better agreement with
those reported by the Q-PCR method.

Figure \ref{Z-scorevsBeta} is the plot of the $Z_i$ versus $E(\beta_i
|\gamma_i=1,x,y)$ for the $86$ shRNAs. The five shRNAs that show
largest discrepancy between the $Z$-score $Z_i$ and posterior activity
change coefficient target five genes: RPS6, KIAA1446, INPP4A,
SFRS1,
and TYK2; this discrepancy is decided by the size of the residues when
a linear regression is fitted. Table \ref{ZscoreLargestD} reports the
posterior probability of activity change, posterior activity change
coefficient, posterior mean of cell viability $E(\mu_i|x,y)$, $Z$-score
$Z_i$, and Q-PCR activity change $k_i$ for these five shRNAs showing
largest discrepancy. All these methods claim unequivocally and
unanimously that the shRNA targeting RPS6 decreases the pathway
activity significantly. In fact, we have verified that RPS6 is a~crucial factor for HCV and we are preparing a manuscript for this
finding. Both Q-PCR and our method claim that the one targeting
KIAA1446 increases the activity of the pathway and that targeting
INPP4A has little effect on the pathway, while the $Z$-score approach
suggests differently; for SFRS1, Q-PCR seems to claim a mild increase
of the pathway activity, our method\vadjust{\goodbreak} suggests little effect on the
pathway activity and $Z$-score suggests a decrease in pathway activity;
for TYK2, Q-PCR suggests little effect and both our method and $Z$-score
suggest a decrease in pathway activity. Table~\ref{ZscoreLargestD}
seems to suggest again that our method is in better agreement with the
Q-PCR method in reporting the activity change effects of shRNAs. We
note that the percentage in the cell viability in Table~\ref
{ZscoreLargestD} shows the corresponding percentile among all the
$6\mbox{,}549$ wells.

%
\begin{table}
\caption{The $Z$-score, posterior activity change coefficient, and Q-PCR
activity change effect $k_i$ for~five~shRNAs}
\label{ZscoreLargestD}
\begin{tabular*}{\tablewidth}{@{\extracolsep{\fill}}lccd{2.3}cd{2.3}c@{}}
\hline
& \multicolumn{1}{c}{\textbf{Gene}} &
\multicolumn{1}{c}{$\bolds{1-p_i}$} & \multicolumn{1}{c}{$\bolds{E(\beta_i |\gamma_i=1,x,y)}$}
& \multicolumn{1}{c}{$\bolds{E(\mu_i|x,y)}$} & \multicolumn{1}{c}{$\bolds{Z_i}$}
& \multicolumn{1}{c@{}}{$\bolds{k_i}$}
\\
\hline
75&RPS6 &0.993 &-2.323 &$-$0.657 (8.80\%)&-3.067 &0.08 \\
66&KIAA1446&0.491 & 0.973 &$-$0.657 (8.81\%)&-0.134 &2.59 \\
28&INPP4A &0.005 &-0.088 &$-$1.065 (6.00\%)&-1.925 &1.25 \\
27&SFRS1 &0.005 &-0.111 &$-$0.975 (6.41\%)&-1.719 &1.83 \\
46&TYK2 &0.907 &-1.542 &$-$0.650 (8.95\%)&-2.317 &1.33 \\
\hline
\end{tabular*}
\end{table}
%

\subsection{Sensitivity analysis}\label{Sensitivityanalysis}
This subsection examines certain distributional assumptions on the
prior used in Section \ref{Analysisandvalidation} for analyzing
the HCV study data. The prior on $p$ used was $\operatorname{\mathcal{B}eta}(9,1)$; we
now also consider $\operatorname{\mathcal{B}eta}(1,1)$. The prior on $\mu_i$ was
$\mathcal{U}(-3,1)$; we now also consider the empirical one
$(0.248+2.77) \cdot\operatorname{\mathcal{B}eta}(6,2)-2.77$. Thus, there are four
combinations in these comparisons, with all the other analysis
strategies unchanged. We report here three posterior inferences for the
comparison of these four studies. In fact, we made extensive
comparisons and found only similar results and the three reported here
are the most representative ones. Table \ref{summarypi} treats the
%
\begin{table}[b]
\caption{Posterior probability of incurring activity change $1-p_i$ for
the $145$ control wells}\label{summarypi}
\begin{tabular*}{\tablewidth}{@{\extracolsep{\fill}}lcccc@{}}
\hline
\textbf{Prior on} $\bolds{p}$\hspace*{-18pt} &\multicolumn{2}{c}{$\bolds{\operatorname{\mathcal{B}eta}(9,1)}$}
& \multicolumn{2}{c@{}}{$\bolds{\operatorname{\mathcal{B}eta}(1,1)}$}\\[1pt]
\ccline{2-3,4-5}\\[-8pt]
\textbf{Prior on} $\bolds{\mu_i}$\hspace*{-18pt} & \multicolumn{1}{c}{\textbf{Empirical prior}}
& \multicolumn{1}{c}{\textbf{Uniform prior}}
& \multicolumn{1}{c}{\textbf{Empirical prior}}
& \multicolumn{1}{c@{}}{\textbf{Uniform prior}}\\
\hline
Median &0.0045 &0.0068 &0.0043 &0.0038\\
75\% quantile &0.0095 &0.0155 &0.0078 &0.0065\\
Maximum &0.1275 &0.2532 &0.1098 &0.0955\\
\hline
\end{tabular*}
\end{table}
posterior probability of activity change based on the $145$ controls.
That all these probabilities in Table \ref{summarypi} are close to
zero shows that all of them perform very well, just as the one in
Section \ref{Analysisandvalidation}. Table \ref{picor} reports
%
\begin{table}
\caption{Correlations between the posterior probability of activity
change obtained from one prior combination and that from
another}\label{picor}
\begin{tabular*}{\tablewidth}{@{\extracolsep{\fill}}lcccc@{}}
\hline
\textbf{Prior on} $\bolds{p}$
& & \multicolumn{2}{c}{$\bolds{\operatorname{\mathcal{B}eta}(9,1)}$}
& \multicolumn{1}{c@{}}{$\bolds{\operatorname{\mathcal{B}eta}(1,1)}$}\\[1pt]
\ccline{3-4,5-5}\\[-8pt]
& \textbf{Prior on} $\bolds{\mu_i}$
& \multicolumn{1}{c}{\textbf{Empirical prior}}
& \multicolumn{1}{c}{\textbf{Uniform prior}}
& \multicolumn{1}{c@{}}{\textbf{Empirical prior}}\\
\hline
$\operatorname{\mathcal{B}eta}(9,1)$ &Empirical prior \\
&Uniform prior &0.972 & & \\
$\operatorname{\mathcal{B}eta}(1,1)$ &Empirical prior&0.997 &0.960 &\\
&Uniform prior &0.992 &0.946 & 0.995\\
\hline
\end{tabular*}
\vspace*{-10pt}
\end{table}
%
%
\begin{table}
\caption{Correlations between the posterior activity change coefficient
obtained from one prior combination and that from
another}\label{betaicor}
\begin{tabular*}{\tablewidth}{@{\extracolsep{\fill}}lcccc@{}}
\hline
\textbf{Prior on} $\bolds{p}$
& & \multicolumn{2}{c}{$\bolds{\operatorname{\mathcal{B}eta}(9,1)}$}
& \multicolumn{1}{c@{}}{$\bolds{\operatorname{\mathcal{B}eta}(1,1)}$}\\[1pt]
\ccline{3-4,5-5}\\[-8pt]
& \textbf{Prior on} $\bolds{\mu_i}$
& \multicolumn{1}{c}{\textbf{Empirical prior}}
& \multicolumn{1}{c}{\textbf{Uniform prior}}
& \multicolumn{1}{c@{}}{\textbf{Empirical prior}}\\
\hline
$\operatorname{\mathcal{B}eta}(9,1)$ &Empirical prior \\
&Uniform prior &0.990  \\
$\operatorname{\mathcal{B}eta}(1,1)$ &Empirical prior&0.992 &0.988 \\
&Uniform prior &0.990 &0.988 & 0.989\\
\hline
\end{tabular*}
\vspace*{-3pt}
\end{table}
the correlation between the probability of activity change obtained
from one combination and that obtained from another. Table \ref
{betaicor}, like Table \ref{picor}, reports the correlation regarding
the posterior activity change coefficient obtained from different
analyses. Since all of the correlations in Table \ref{picor} and Table
\ref{betaicor} are larger than $0.94$, it seems that the posterior
inferences are not sensitive to these aspects of the prior assumptions.

\subsection{Some comparison with the literature}\label{Somecomparison}
\citet{C4} reviewed five papers that use siRNA screens to identify
cellular factors that impact replication of HCV or subgenomic replicons
and found little overlap between them. As part of our effort to
evaluate the performance of our method, we select all the siRNAs in our
list that show activity change in \citet{S2} and \citet{T1} and see whether our findings are
in line with theirs. A
comprehensive discussion of the scientific findings of our study will
be reported elsewhere.

\citet{S2} reported that siRNAs specific for three human
kinases, CSK, JAK1, and VRK1, were identified to reduce the replication
of HCV; they also reported that by examining the siRNA knockdown effect
of the eight kinase genes in the Src family (BLK, HCK, FGR, LCK,
LYN,
FYN, c-SRC, and YES) on the HCV replicon, they found that their siRNAs
targeting seven of them did not show any effect and that targeting FYN
elevated replicon levels by about 3-fold upon transduction.

In our HCV study, there are shRNAs targeting nine of these $11$ genes,
except c-SRC and YES. We now compare our results on these nine genes
with those from \citet{S2}.\vadjust{\goodbreak}

Among the five shRNAs targeting FYN, four of them indicated
considerable elevation in HCV replication and the other one did not
show activity change; none showed change in cell viability. Our
screening showed also that none of the shRNA targeting FGR or HCK
indicated any appreciable activity change. These results are in perfect
or excellent agreement with those in \citet{S2}.

Among the four targeting BLK, three of them showed no activity change
and one indicated considerable increase; among the five targeting LCK,
four of them showed no activity change and one indicated considerable
decrease. These are in partial agreement with \citet{S2}.

Among the $10$ shRNAs targeting CSK, one showed considerable decrease
both in HCV replication and in cell viability; another showed
considerable decrease in HCV replication without change in cell
viability; still another showed considerable increase in HCV
replication without change in cell via\-bility; the remaining seven
showed no change. All the four targeting JAK1 indicated no activity
change and no cell viability change. None of the two targeting VRK1
indicated any appreciable activity change, although one of them
indicated considerable reduction in cell viability. Among the five
shRNAs targeting LYN, three of them indicated strong elevation in HCV
replication without appreciable change in cell viability. These suggest
a certain amount of disagreement between our findings and those in
\citet{S2}.

In view of the general lack of concordance between results of similar
HCV screens in the literature, the amount of agreement between \citet{S2} and our study seems encouraging.

We note that \citet{T1} reported that there is little overlap
between the results of their screen and those of \citet{S2}; in fact, none of the three genes CSK, JAK1, and VKR1 were
selected in the primary screening of \citet{T1}, which seems to
be in agreement with ours.

We now compare our results with \citet{T1}, in which $96$ genes
are chosen from the primary screening for reduced HCV replication. In
the shRNA list of our study, there are shRNAs targeting four of these
$96$ genes; they are TBK1, NUAK2, COASY, and MAP3K14.

Two of the four shRNAs in our list targeting MAP3K14 showed strong
reduction in HCV replication, another indicated marginal reduction in
HCV replication and the fourth one indicated no activity change; none
showed change in cell viability. The shRNAs targeting the remaining
three genes showed little effect on the reduction of HCV replication.
We note that \citet{T1} used $21\mbox{,}094$ siRNA pools targeting the
entire human RefSeq transcript database; each of these pools consists
of four individual siRNA duplexes and each of these four siRNAs targets
a different sequence within the target transcript.

\section{Discussion}\label{DISCUSSION}

We have presented a Bayesian method to analyze data from two channel
cell-based RNAi HTS experiments with replicates, in which the phenotype
of a pathway-specific reporter gene and that of a constitutive reporter
are measured. These experiments are typical in screens for signaling
pathway components and the purpose is often to identify genes that
affect the activity of the specific pathway without affecting that of
the constitutive reporter.

We have conducted simulation studies and real data analysis to
illustrate the methods. Our simulation studies indicate that error
terms with shRNA specific $t$-distribution do make the method flexible
and robust and that replication provides better power in identifying
shRNAs of interests and, at the same time, gives better estimates of
the false discovery rates.

In our analysis of the real data set, we illustrate the usage of
negative controls, included originally for normalization purposes only,
to assess the performance of our methods, to estimate false discovery
rate, and to prioritize the shRNAs for secondary validation, in
addition to hit selection; we find our methods perform excellently and
these negative controls are very useful. We have also conducted a Q-PCR
based assay to assess the activity change of $86$ shRNAs; we find the
results based on Q-PCR are in better agreement with those based on our
method than with those based on the standard $Z$-score approach. We have
also shown that our method is insensitive to the choice of prior.
Finally, it is encouraging to find that there does exist some overlap
between the results of the HCV study and those of \citet{S2} and \citet{T1}.

Realizing the usefulness of negative controls, we are interested in
knowing the optimal number of negative controls to be included in an
experiment. It seems also desirable to include positive controls as
well in the experiment. With both positive and negative controls, we
may extend our Bayesian method so as to improve our report on the false
positive and negative rates in a RNAi HTS. We note that characteristics
of these controls, especially positive controls, and the number of
these controls deserve serious attention in designing an experiment and
in extending our statistical method.

%

\section*{Acknowledgments}

We thank the referees, the Associate Editor, and the Area Editor for
constructive criticisms that led to significant improvement of the
model and the presentation in this paper.


\begin{supplement}
\stitle{A Computer algorithm for analyzing data from two-channel
cell-based RNAi experiments with replicates}
\slink[doi]{10.1214/11-AOAS496SUPP}
\slink[url]{http://lib.stat.cmu.edu/aoas/496/supplement.pdf}
\sdatatype{.pdf}
\sdescription{This note provides the hybrid MCMC algorithm for sampling
the posterior distribution used in \citet{C3} and several
observations used in designing this algorithm so as to make it more
efficient.}
\end{supplement}

%

\printaddresses

\end{document}